\documentclass[aps,prx,superscriptaddress,twocolumn,showpacs,longbibliography]{revtex4-1}
\usepackage{amsmath, amsthm, amssymb}
\usepackage{bbold}
\usepackage{graphicx}
\usepackage{color}
\usepackage{times, verbatim}

\newcommand{\ket}[1]{{\left\vert{#1}\right\rangle}}

\usepackage{natbib}
\usepackage{hyperref}
\hypersetup{
        colorlinks=true,
}

\newcommand{\eqnref}[1]{(\ref{#1})}

\newcommand{\Tt}{T_\text{tot}}

\begin{document}

\title{Analyzing many-body localization with a quantum computer}

\author{Bela Bauer}
\affiliation{Station Q, Microsoft Research, Santa Barbara, CA 93106, USA}

\author{Chetan Nayak}
\affiliation{Station Q, Microsoft Research, Santa Barbara, CA 93106, USA}
\affiliation{Physics Department, University of California,  Santa Barbara, CA 93106, USA}

\begin{abstract}
Many-body localization, the persistence against electron-electron interactions of the localization of states with non-zero excitation energy density, poses a challenge to current methods of theoretical and numerical analysis.
Numerical simulations have so far been limited to a small number of sites,
making it difficult to obtain reliable statements about the thermodynamic limit.
In this paper, we explore the ways in which a relatively small quantum computer could be
leveraged to study many-body localization. We show that, in addition to
studying time-evolution, a quantum computer can, in polynomial time, obtain
eigenstates at arbitrary energies to sufficient accuracy that localization can be observed.
The limitations of quantum measurement, which preclude the possibility of directly obtaining the
entanglement entropy, make it difficult to apply some of the definitions of many-body
localization used in the recent literature. We discuss alternative tests of localization that can be implemented on a quantum computer.
\end{abstract}

\maketitle

\section{Introduction}

Since the seminal contributions of Gornyi {\it et al}~\cite{Gornyi2005} and Basko {\it et al.}~\cite{Basko06a,Basko06b},
the question of whether Anderson localization
can persist against interactions at non-zero excitation energy density has been revisited.
The possible existence of such a phenomenon, dubbed many-body localization (MBL),
is closely intertwined with other open fundamental
questions in quantum statistical mechanics: whether the eigenstate thermalization hypothesis~\cite{Deutsch91,Srednicki94}
holds for generic quantum systems and whether isolated quantum systems
can equilibrate \cite{Rigol08}. In studying these questions, significant theoretical and numerical evidence~\cite{Oganesyan07,DeChiara2006,Znidaric08,Pal2010,Bardarson12,Iyer12,Vosk13,Serbyn2013,serbyn2013-1,Huse2013,bahri2013,pekker2013,yao2013,chandran2013,bauer2013,Kjaell14,Serbyn14}
has been assembled indicating that
a many-body localized phase exists; for a recent review, see Ref.~\onlinecite{nandkishore2014review}.
More recently, the existence of a many-body localized phase has been proven rigorously
in a class of spin chains~\cite{imbrie2014}.

However, the numerical simulation of putative many-body localized systems
has remained extremely challenging: while these states have low
entanglement and should be well-approximated by tensor-network states~\cite{bauer2013},
the known methods to find such states are best-suited for finding ground states,
and become inefficient when targeting states at generic energies in the spectrum,
where the gaps to nearby states are exponentially small in the system size.
Another approach has been to study the time evolution of such systems from
easily-prepared initial states, which is limited to small systems by the unbounded growth
of entanglement \cite{Bardarson12}.
This has limited accurate computations of the properties of many-body
localization to systems of approximately 20 sites or smaller.

In this paper, we ask how a quantum computer of moderate size
could be used to break through this barrier. Quantum computers are
naturally suited to simulating the dynamics of quantum systems;
indeed, the idea was first conceived in this context~\cite{feynman1982}.
However, the devil is in the details: on a ``digital'' quantum computer
operating within the gate model~\cite{Deutsch1989}, the evolution of a quantum
system is mapped to a series of one- and two-qubit gates chosen from a finite,
but computationally universal, set. This mapping can induce enormous overhead in practice.
Moreover, one does not have access to the full resulting quantum wavefunction but, rather, to information that can be extracted from projective measurements of individual qubits,
although one has the freedom to choose the basis in which these measurements are done.
Recently, it has become evident that while there are
quantum algorithms that have exponential speedup over their classical counterparts,
many of these algorithms have such an advantage only in an
asymptotic regime that is unlikely to be reached on a quantum computer in the foreseeable future, even without even taking into account additional overhead necessitated by
quantum error correction.
Consider two examples: Shor's celebrated algorithm for factoring~\cite{Shor1994,shor1997polynomial}
and the simulation of quantum chemistry~\cite{yung2012introduction}. In the first case, a quantum
computer would require thousands of qubits to factor a number
that cannot be factored using the most efficient classical algorithms. In the second
case, a moderate number of qubits is required, but the mapping of the
unitary evolution into the gate model requires
the coherent execution of a very large number of gates~\cite{wecker2013,hastings2014,poulin2014}.

In this paper, we will discuss how the power of a quantum computer
can be brought to bear on understanding the phenomenon of many-body localization.
We demonstrate that even a relatively small quantum computer will have significant advantages
over a classical computer in analyzing Hamiltonians that may exhibit many-body localization.

In doing so, we address another important question: are the hypothesized properties of
MBL eigenstates observable in experiments?
An exact eigenstate of the Hamiltonian with non-zero excitation energy density
cannot be prepared efficiently -- i.e. in a time that scales only polynomially
with system size -- since the time required
depends inversely on the required energy resolution and the energy level spacing
in the middle of the spectrum is exponentially small in the system
size.~\footnote{This may not apply if the
relevant energy scale is a local scale instead of a global one.} Even though
our main interest in this paper is \emph{not} analog \emph{simulation} but, rather,
quantum \emph{computation} of the properties of MBL states with a general purpose
quantum computer, it is useful to momentarily regard a quantum computer
as an experimental system, albeit a very idealized one. Then, our results
demonstrate that the characteristic properties of exact
energy eigenstates can, indeed, be observed
in approximate eigenstates that can be
prepared in polynomial time. Hence these properties are likely
also observable in other, less idealized, experimental setups.

In Section~\ref{sec:preparation}, 
we first review briefly how the time evolution of a quantum system is mapped into the gate
model using a Trotter-Suzuki decomposition, since this is a basic building block
for all that follows. We show that typical features of many-body localized Hamiltonians,
such as short-ranged interactions and on-site disorder, make this decomposition
feasible for system sizes of interest. A quantum computer
can thereby determine the time evolution of an easily-prepared initial state
such as a random product state. This may be viewed as a computation of
the {\it global quench dynamics} of the system where the system is, at least
initially, very far away from equilibrium. Such a computation would allow us to probe the
equilibration and thermalization properties of the system, which can reflect
its localization properties.

We then explore how random {\it energy eigenstates} of the system can be
prepared with sufficient accuracy to observe signatures of
many-body localization with a polynomial-depth quantum circuit.
This is done using the quantum phase estimation algorithm. However,
unlike in many applications in which one is interested in finding ground states,
a generic eigenstate that results from quantum phase estimation is relevant to the
study of many-body localization.
This key step of preparing eigenstates greatly enhances the usefulness of a
quantum computer since it opens up the study of situations other than global quenches,
and in particular gives access to the dynamical response of the system to weak perturbations.
Examples include transport measurements or local quenches in systems prepared at fixed energy.
This may also allow more quantitative connections to experiments, such as states of
ultra-cold atoms in optical traps.

However, preparing a state -- either through quench dynamics or through quantum phase
estimation -- is only half the battle. We are now faced with extracting information from this state.
In contrast to classical simulations,
we cannot simply examine the wavefunction directly and deduce all of its properties.
In particular, measuring the entanglement entropy in the state is difficult, if not impossible, within
the constraints of our setup. Instead,
we are limited to performing unitary operations on the state and then performing local,
projective measurements, 
thereby obtaining a string of zeroes and ones -- as many classical bits of information as measured qubits.
The question thus arises how we can characterize the eigenstates.
In Section~\ref{sec:meas}, we discuss in more details the limitations of the measurement process
and propose scenarios how the eigenstates can be characterized, where we
focus mostly on measurements of transport properties at finite energy densities. The setups we consider
are \emph{local} quench dynamics, as well as probing the response of the system to a ``tilt''.
Finally, we analyze the possible effects of errors, such as decoherence and discretization errors, on the computation
of MBL states and their properties.

\section{State preparation} \label{sec:preparation}

\subsection{Models, Representation, and Time Evolution}

The models that we focus on here are (1) a model of spinless fermions
with nearest-neighbor hopping in one dimension, nearest-neighbor interactions, and on-site disorder
and (2) an XXZ spin chain with a random Zeeman field in the $z$-direction.
The first model has Hamiltonian:
\begin{equation} \label{eqn:H-fermions} \begin{split}
H_{\rm f} = -t \sum_{i=1}^{L-1} \left( c_i^\dagger c_{i+1} +c^\dagger_{i+1}c_i \right) \\ + \sum_{i=1}^L w_i n_i + V\sum_{i=1}^{L-1} n_i n_{i+1} ,
\end{split} \end{equation}
where $c^\dagger_i$ creates a spinless fermion on site $i$, and $n_i=c^\dagger_i c_i$. The $w_i$ are uniformly chosen from $w_i \in [-W,W]$.
The second has:
\begin{equation} \label{eqn:H-spins} \begin{split}
H_{\rm s} = -J_\perp \sum_{i=1}^{L-1} \left( S^x_i S^x_{i+1} + S^y_i S^y_{i+1} \right) \\
+ \sum_{i=1}^L w_i S^z_i/2 +J_z  \sum_{i=1}^{L-1} S^z_i S^z_{i+1}.
\end{split} \end{equation}
For open boundary conditions, these models have the same spectrum for $J_\perp = t$ and $V=2J_z$,
while for closed (i.e., periodic or antiperiodic) boundary conditions, more care must be taken when
relating them to each other through a Jordan-Wigner transformation.

We would like to compute time evolution due
to these Hamiltonians on a general-purpose quantum
computer operating under the circuit model.
Therefore, the unitary evolution $U = \exp(-i T H)$ must be mapped
to a series of gates chosen from a given set of available gates. We will call this procedure \emph{compiling} below, in analogy
to the well-known classical procedure of compiling a program in a high-level programming language into the assembly code,
i.e. machine language, of the target hardware platform.
While many approaches to achieve
this are known (for some recent improvements, see e.g. Refs.~\cite{berry2012black,anmer2012,berry2013exponential2}),
by far the most widely-used is the Trotter-Suzuki
decomposition~\cite{trotter1959,suzuki1976}. First, the time evolution is broken up into a series of time
steps $\delta t = T/N$:
\begin{equation}
U = e^{-i \delta t H} \ldots e^{-i \delta t H}
\end{equation}
Then we write $H=\sum_{i=1}^{m} H_i$, where the $H_i$ are chosen such
that $U_i(\delta t) = \exp(-i \delta t H_i)$ can be compiled into a series of gates exactly.
Then, if we use a first-order Trotter-Suzuki decomposition, we write
\begin{equation}
e^{ -i \delta t H } = \left(\prod_{j=1}^m e^{-i \delta t H_j }\right) + O({\delta t^2}).
\end{equation}
This decomposition is only accurate to order $\delta t^2$, but
a more elaborate decomposition can be found that is accurate to any given order in $\delta t$.
However, the number of terms in the decomposition grows quickly with the order.
Therefore, the optimal order depends on the desired
accuracy $\epsilon$ (in trace-norm distance on the final state), the total time $T$,
and the norm of the Hamiltonian operator.
Ref.~\onlinecite{efficient2007} gives both an upper bound on the total number $N_\mathrm{exp}$
of separate exponentials $U_i(\delta t) = \exp( -i \delta t H_i)$ that must be executed to achieve
a given accuracy in trace norm distance, as well as an estimate for the optimal
order of Trotter decomposition.
In order to determine the number of elementary gates required to perform the evolution $U$,
we multiply $N_\text{exp}$ by the number of elementary gates needed to
perform each of the $U_i(\delta t)$.

In the models that are relevant to many-body localization, both kinetic and interaction terms
are generally local. Consider, for example, the Hamiltonian given in Eqn.~\eqnref{eqn:H-fermions}.
It can be expressed as a sum of $3$ non-commuting terms:
\begin{eqnarray}
\label{eqn:H-fermions-3terms}
H_1 &=& -t \sum_{i=1}^{(L-2)/2} \left( c_{2i}^\dagger c_{2i+1} +c^\dagger_{2i+1}c_{2i} \right) \nonumber \\
H_2 &=& -t \sum_{i=1}^{L/2} \left( c_{2i}^\dagger c_{2i-1} +c^\dagger_{2i-1}c_{2i} \right) \\
H_3 &=&  \sum_{i=1}^L w_i n_i + V\sum_{i=1}^{L-1} n_i n_{i+1}, \nonumber
\end{eqnarray}
Here, we have taken $L$ to be even.
More generally, the number of non-commuting terms is
$m = 1+z$ on a regular lattice in $d$ dimensions with coordination number $z$,
where 1 accounts for the diagonal terms (interaction and on-site potential), and
$z$ accounts for the kinetic terms.
Crucially, this is independent of the size of the system;
however, $\| H \|$ is extensive~\footnote{Note that in Ref.~\onlinecite{wecker2013}, it was pointed
out that the bound of Ref.~\onlinecite{efficient2007} can be made more tight and stated in
terms of the norm of the individual non-commuting terms;
however, in our setup the norm of these is also extensive and this tighter bound makes no
difference.}, thus the overall scaling is slightly faster than $L \cdot T$.

The individual terms in the Trotter expansion of $U$ still need to be expressed in the
available gate set. At this stage, the fermionic (\ref{eqn:H-fermions})
and spin (\ref{eqn:H-spins}) Hamiltonians are equivalent,
so we use terminology appropriate to the latter. We write
$H_i = \sum_j h_i^j$, where $h^i_j$ is a product of Pauli matrices, and use
\begin{equation}
\exp \left( -i \delta t H_i \right) = \prod_j \exp \left(-i \delta t h_i^j \right).
\end{equation}
These gates can be transformed into a basic gate set, see Refs.~\onlinecite{whitfield2011,wecker2013}
for details. We find the following parallel gate counts, i.e. assuming that gates that operate on
different qubits can be executed simultaneously: (i) For $H_1$ and $H_2$, we need 10 gates each.
(ii) For $H_3$, we again need $1+z$ gates, where $z$ is the coordination number of the lattice.
If the original Hamiltonian is fermionic then,
for dimensions $d>1$, there will be additional Jordan-Wigner strings, but
their overhead can be greatly reduced~\cite{bravyi2002,CodyJones2012,hastings2014}.

For how much time $T$ must we we evolve the system to observe physical
manifestations of many-body localization?
It has been shown that within the scenario of a global quench, certain properties,
such as the saturation of entanglement entropy to a volume law~\cite{Bardarson12},
can be observed only after time that is exponentially-large in the system size. This makes it experimentally
unfeasible to observe these properties and ultimately renders them unphysical. A more
appealing scenario may be that of a local quench~\cite{Kjaell14}. In the MBL phase, the perturbation
only propagates a finite distance and the long-time behavior is observed after a time that does not
scale with system size.

As the numerical studies described in the following sections show, the Trotter error need not
be kept extremely small for the purpose of detecting MBL physics; instead, other sources
of error, such as limitations on the time to which quantum phase estimation can be run,
are more relevant. If, however, a quench or transport scenario requires very low error
bounds on the unitary evolution, the methods recently put forward in
Ref.~\onlinecite{berry2013exponential2} may be favorable over a Trotter decomposition.

\subsection{Quantum phase estimation}

\begin{figure}
  \includegraphics{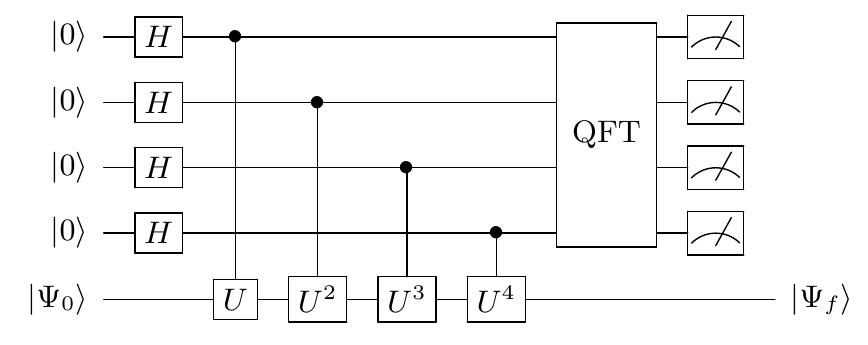}
  \caption{Overview of the quantum phase estimation algorithm discussed in the main text
  for $k=4$ ancilla qubits. The lowest line in the figure indicates the $N$ qubits used for the
  physical system. Here, QFT denotes the quantum Fourier transformation. After the the
  measurements, the readout on the $k$ ancilla qubits contains an estimate for the energy,
  whereas the $N$ qubits for the physical system contain
  the final state $|\Psi_f\rangle$ of Eqn.~\eqnref{eqn:psif}, which is an approximation
  to an eigenstate and can processed further to obtain measurements on the physical system.
  \label{fig:overview} }
\end{figure}

In order to make our discussion self-contained, we briefly review some essential features
of the quantum phase estimation algorithm; see also Fig.~\ref{fig:overview}. More detailed pedagogical discussions can
be found in, for instance, Refs.~\onlinecite{kitaev1995,kitaev2002book,nielsen2010quantum}. Let us suppose that we would like
to find an eigenvalue and corresponding eigenvector of a unitary operator $U$ acting
on a Hilbert space of dimension $2^N$. For us, the unitary operator will be the exponential
of a Hamiltonian, $U=e^{-i T H}$, that we wish to test for many-body localization.
To perform quantum phase estimation, we consider a system of $N+k$ qubits,
where we refer to the first $N$ qubits as the data qubits, on which the operator $U$ acts,
and the next $k$ qubits as ancillas. We will assume that
we can perform Hadamard gates, controlled-$U$ gates, and
arbitrary controlled-phase gates. Controlled-$U^k$ gates can be implemented by applying
the controlled-$U$ gate $k$-times in succession. Suppose that our ancillas are all initially in the state
$|0\rangle$ and the $N$ data qubits are in an arbitrary initial state $|{\psi_0}\rangle$, chosen at random.
Then, we perform Hadamard gates on each of the ancillas,
thereby putting each in the state $(|0\rangle + |1\rangle)/\sqrt{2}$. We then act with a controlled-$U$ gate in which the first ancilla is the control and the $N$ data qubits
are the target on which the unitary acts when the ancilla is in the state $|1\rangle$.
We act with a controlled-$U^2$ gate in which the second ancilla is the control and the data qubits
are the target. We continue similarly with each ancilla so that the $j^{\rm th}$-ancilla
is the control for a controlled-$U^{2^{j-1}}$. The resulting state is
$$
\sum_{\{{i_n} = 0,1\}} U^{{i_1}+2{i_2}+\ldots+2^{k-1}i_{k}}|{\psi_0}\rangle\otimes |{i_1},{i_2},\ldots,{i_k}\rangle
$$
If $T$ is the integer whose binary expansion is ${i_1}{i_2}\ldots i_{k}$, then this can be written
in the form:
\begin{equation}
\sum_{T=0}^{2^k -1} U^{T}|{\psi_0}\rangle\otimes |T\rangle
\label{qpe-step1}
\end{equation}
Expanding the initial state of the data qubits in terms of
the eigenstates of $U$, $|\psi_0\rangle = \sum_n {c_n} |n\rangle$,
where $|n\rangle$ has eigenvalue $e^{i E_n}$, we can rewrite
Eq.~(\ref{qpe-step1}) in the form
\begin{equation}
\sum_n \left[{c_n} |n\rangle \otimes \Biggl(\sum_{T=0}^{2^k -1} e^{i{E_n}T}|T\rangle\Biggr)\right]
\label{qpe-step2}
\end{equation}
We now apply the (inverse) quantum Fourier transform on the ancillas, which acts
on a basis vector according to:
\begin{equation}
|T\rangle \rightarrow \sum_{J=0}^{2^k -1} e^{-2\pi i T\frac{J}{2^k}} |J\rangle
\end{equation}
This results in the state
\begin{equation}
\sum_n \left[{c_n} |n\rangle \otimes \Biggl(\sum_{J=0}^{2^k -1}
g({E_n}-\mbox{$\frac{2\pi J}{2^k}$})
|J\rangle\Biggr)\right]
\label{qpe-step3}
\end{equation}
where $g(x)=(1-e^{- i{2^k}x})/(1-e^{-ix})$. The function $g(x)$ is peaked around
$x=0$. If we increase $k$ so that ${2^k}{E_n}/2\pi$ approaches an integer, then
it becomes more strongly peaked. If ${E_n}/2\pi$ is a $k$-bit binary number modulo
integers, then ${2^k}{E_n}/2\pi$ is an integer and $g(x)=\delta_{{2^k}x,0}$. Then we have
\begin{equation}
\sum_n \left[{c_n} |n\rangle \otimes
|{2^k}{E_n}/2\pi\rangle\Biggr)\right]
\label{qpe-step4}
\end{equation}
By measuring the ancilla, we obtain ${2^k}{E_n}/2\pi$ with probability $|c_n|^2$
and the data qubits are left in the state $|n\rangle$. While the eigenvalue
$E_n$ is the primary goal for applications to period-finding, our main goal here is
to obtain the state $|{\psi_n}\rangle$. Moreover, one is often interested in
finding the ground state of a Hamiltonian and, therefore, needs to choose an initial
state $|{\psi_0}\rangle$ with high overlap with the ground state, so that $|{c_0}|^2$
is not too small. In the present application to many-body localization, however, we are
interested in generic states, so we can take a random initial state.

If ${2^k}{E_n}/2\pi$ is not an integer, then when we measure the ancillas,
we obtain an approximate eigenvalue $E_{\rm approx}$
that is $2\pi/2^k$ times a $k$-bit integer. The data qubits are in the state
\begin{equation} \label{eqn:psif}
|\Psi_f\rangle = \sum_n {c_n} g({E_n}-{E}_{\rm approx}) |n\rangle
\end{equation}
which is not an energy eigenstate, but has amplitude that is sharply peaked at eigenstates
that are near ${E}_{\rm approx}$. We can make it more sharply peaked, thereby obtaining an eigenstate to within any desired accuracy, by increasing the number of ancillas.

In practice, it may be favorable to use an iterative quantum phase estimation (IQPE) algorithm, as described
in Ref.~\onlinecite{parker2000}, which performs effectively the same calculation as outlined above
but requires only one ancilla qubit. This is particularly useful in the context of classical
simulation of the quantum computer for validation purposes, as the classical simulation time
exponentially in the number of qubits unless an approximate method is used.

It remains to be confirmed that
(i) we obtain all states with sufficient probability, and
(ii) we can prepare these states to sufficient accuracy to observe signatures of many-body
localization with a total computation time that scales at most polynomially in the system
size $L$ even near the middle of the spectrum, where gaps to adjacent states are exponentially
small in $L$.

Other approaches of obtaining eigenstates on a quantum computer seem possible: for example,
one could attempt to adiabatically cool towards the ground state of $A=(H-\lambda \mathbb{1})^2$,
for some $\lambda \in [-\|H\|, +\|H\|]$, or adiabatically move from an eigenstate of the diagonal
part of the Hamiltonian, which can easily be prepared, to an eigenstate of the full Hamiltonian.
However, these approaches have major drawbacks: In the first approach, the evolution must be
performed under a non-local Hamiltonian. Furthermore, in both cases the accuracy depends
on whether the adiabatic evolution is performed slow enough compared to the relevant
energy scale, which is hard to control.

\subsection{Gate counts}

We now test the procedure outlined above in a numerical simulation. Our goals are (i) to confirm that
we sample from the correct distribution of eigenstates, (ii) to determine the number of gates that
need to be executed to obtain an eigenstate with a given energy standard deviation,
and (iii) to confirm that we can obtain eigenstates to
sufficient accuracy to be able to probe signatures of many-body localization. The last point
will be deferred to the next section.

In our numerical simulations, we study the Hamiltonian~\eqnref{eqn:H-fermions} on an open
chain of $L$ sites.
We perform iterative quantum phase estimation (IQPE) on $U=\exp \left( -i H T \right)$.
To keep eigenvalues from wrapping around the unit circle, we need to ensure that
$T \| H \| < 1$ and thus set $T = [L(2+V+W)]^{-1} < \| H \|^{-1}$. For small systems,
we perform the unitary evolution exactly; for larger systems and to assess the effect of
Trotter errors, we perform a first-order Trotter decomposition. We find that for the system
sizes and evolution times considered here, the Trotter error is not very significant; for larger
systems, a higher-order Trotter decomposition may be favorable. In this setup, the number of
gates (assuming parallel execution of non-overlapping gates) necessary to evolve the system to
a time $\tau$ is given by
\begin{equation} \label{eqn:gc}
N_g = 23\ \tau/\delta t.
\end{equation}
Here, 23 is how many gates are needed to execute $\exp(-i \delta t H_1) \exp(-i \delta t H_2) \exp(-i \delta t H_3)$,
see Eqn.~\eqnref{eqn:H-fermions-3terms}.

To obtain $k$ bits of the desired eigenvalue, we need to evolve for a total time
\begin{equation}
\Tt = T \sum_{i=0}^{k} 2^i = T (2^{k+1}-1).
\end{equation}
This total time effectively determines the absolute accuracy as well as computation cost of IQPE;
the same accuracy can in principle be achieved by reducing $T$ and accordingly increasing $k$,
or vice versa. We therefore plot all results against $\Tt$.

\begin{figure}
  \includegraphics{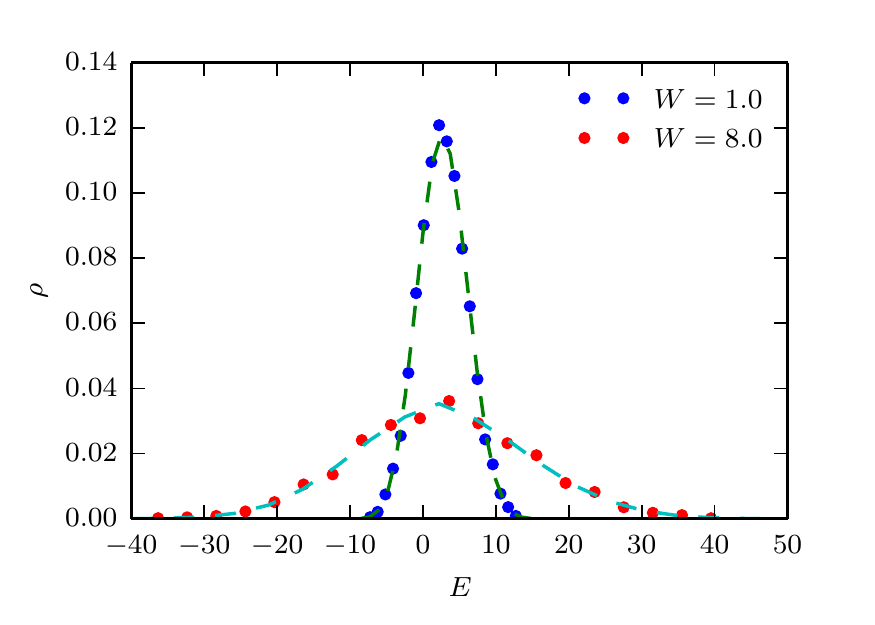}
  \caption{Dashed lines: density of states $\rho(E)$ obtained using full diagonalization. Points: $\rho$ obtained using IQPE for $k=16$ bits. All results are for $L=12$. \label{fig:dos}}
\end{figure}

We first confirm that we obtain states with the correct probability when starting from random initial product states. To this end, we compare the density of states $\rho(E)$ obtained using the IQPE procedure outlined above to that obtained from an exact, full diagonalization of the same Hamiltonians. We consider at least 100 disorder realizations and average over a total of 10000 states. Our results are shown in Fig.~\ref{fig:dos}. Clearly, the agreement between the two approaches is excellent both in the delocalized ($W=1$) and localized ($W=8$) regime.

\begin{figure}
  \includegraphics{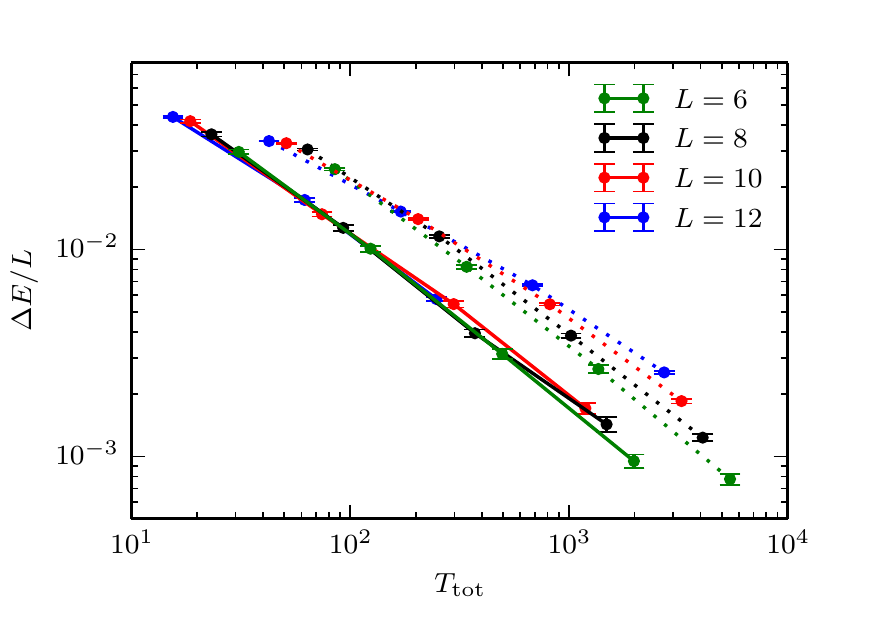}
  \caption{(Color online) Energy standard deviation density $\Delta E/L$.
  Dotted lines are $W=1$, solid lines are $W=8$. Results have been obtained using exact time evolution to
  exclude any Trotter errors. \label{fig:var} }
\end{figure}

As a simple measure of how accurately we can prepare eigenstates, we calculate the energy standard deviation
\begin{equation}
\Delta E = \left\langle \sqrt{\langle H^2 \rangle - \langle H \rangle^2} \right\rangle,
\end{equation}
where the outer average is over output states from different runs of IQPE for different disorder
realizations and initial states.
A naive expectation is that $\Delta E/L \cdot \Tt \sim 1$. To check this, we perform numerical simulations,
again averaging over 10000 states for each choice of $\Tt$, $W$ and $L$. In these simulations, we perform
the time evolution exactly to separate out the effect of Trotter errors. This limits the system size we can
study to $L=12$ because we exponentiate the Hamiltonian exactly. As shown in Fig.~\ref{fig:var}, we find
$\Delta E/L = c \Tt^{-\alpha}$, where $c$ and $\alpha$ are fit parameters. The best fit is obtained for
$\alpha \approx 0.8$, which deviates slightly from the naive expectation of $\alpha=1$. We observe that
all curves for different $L$, but equal $W$ collapse, indicating that $c$ depends only on $W$. We observe
that in the localized regime, $\Delta E/L$ of the final states is lower than in the delocalized regime.

\begin{figure}
  \includegraphics{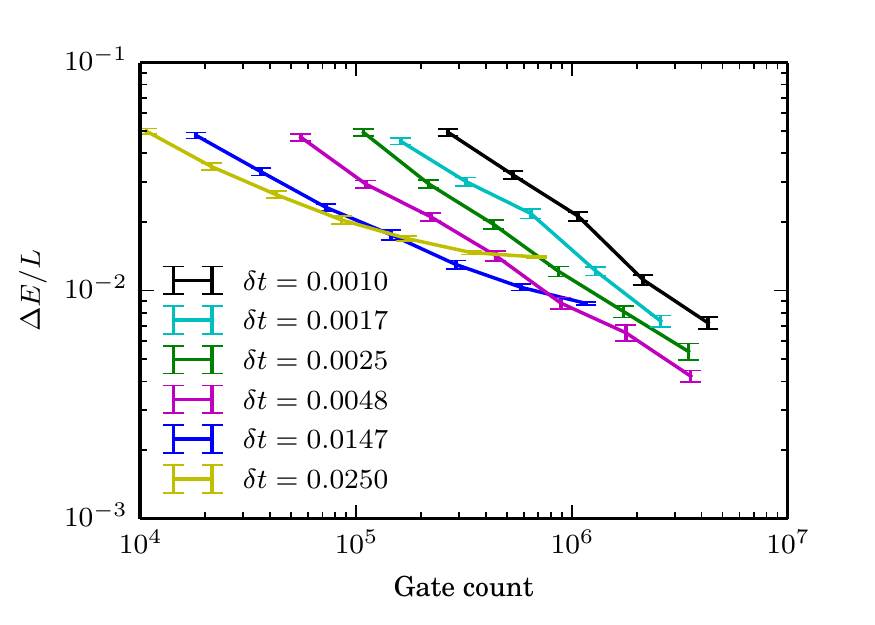}
  \caption{(Color online) Trotter errors for $L=16$, averaging over 1000 instances. Here, we use a first-order
  Trotter decomposition. \label{fig:gc} }
\end{figure}

At this point, we can also analyze Trotter errors and, using Eqn.~\eqnref{eqn:gc},
obtain the gate count necessary to obtain a final state
with some fixed $\Delta E/L$. For this analysis, we restrict ourselves to a first-order
Trotter decomposition. In Fig.~\ref{fig:gc}, we show $\Delta E/L$ against the gate count for
different choices of the Trotter time step $\delta t$.
We observe that the timestep has to be decreased roughly as $\delta t \sim \epsilon$, where $\epsilon$
is the desired $\Delta E/L$. For example, to achieve $\Delta E/L < \epsilon = 0.01$, a timestep
$\delta t = 0.025$ seems necessary; to achieve $\epsilon = 0.005$, $\delta t < 0.0147$ is necessary.

We note that the total gate counts shown here are on the order of a few millions
and thus much more realistic than the gate counts obtained for quantum chemistry
in Ref.~\onlinecite{wecker2013}. Assuming a logical gate rate of $1\ \mathrm{MHz}$, the prepration of an
approximate eigenstate would require a coherence time on the order of 1 second.

\subsection{Observation of MBL}

Having established the gate counts required to reach a certain $\Delta E/L$, the question arises what
$\Delta E/L$ must be achieved in order to observe signatures of many-body localization. Naively, one might expect
that $\Delta E$ must be small compared to the mean level spacing $\delta E$. As the latter is exponentially small in
the system size, this would imply an exponential scaling. In the following,
however, we will argue that, in the many-body localized phase,
there are states that can be prepared in polynomial (in system size) time that
display key signatures of many-body localization.
A similar conclusion was reached from a very different perspective in Refs.~\onlinecite{Nandkishore14a,Johri14}
As a test of the localization properties of the final states obtained in our quantum algorithm, we use the
entanglement entropy. The presence of an area law for the entanglement entropy
has been established previously as a good indicator
of many-body localization in exact eigenstates~\cite{bauer2013}.

It is intuitively clear that even in a many-body localized phase, where exact eigenstates obey an area law,
approximate eigenstates with $\Delta E/L > \delta E$, but small compared to other scales
in the problem, may display very different entanglement properties. A superposition with random coefficients of exponentially
(in the system size) many eigenstates in a given, small energy window will likely
have volume-law entanglement. On the other hand, a superposition of a few eigenstates
that are far away in energy can have the same $\Delta E/L$, but may still display an area law.

\begin{figure}
  \includegraphics{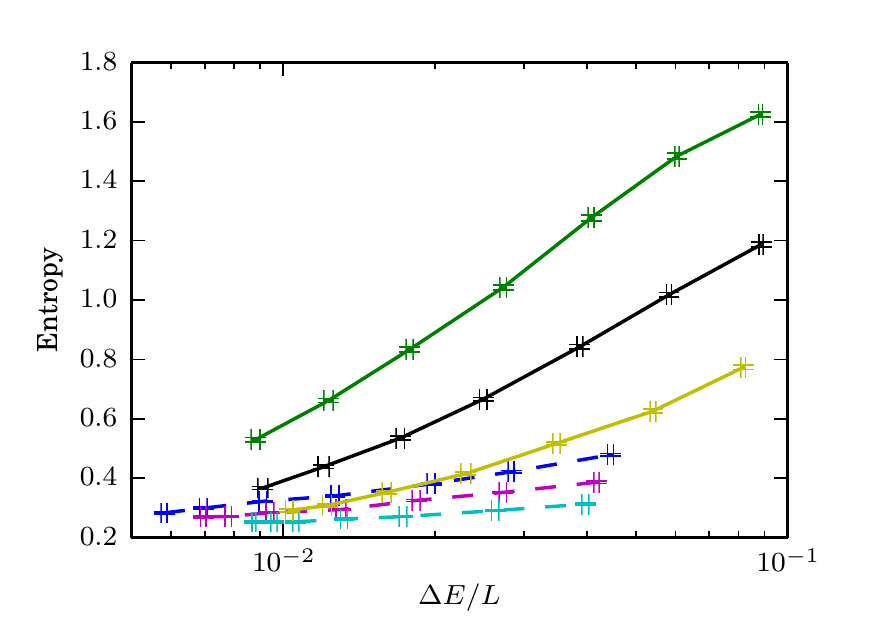}
  \caption{(Color online) The entanglement entropy at the center of the system as a function of $\Delta E/L$ for various system sizes and initial states used in the IQPE algorithm. Dashed lines show data for initial product states in the $S^z$ basis, while solid lines show data for initial states with a mix of $S^z$ and $S^x$ basis, as explained in the text. System sizes are, from top to bottom, $L=12, 10, 8$.
\label{fig:initstate} }
\end{figure}

To explore this quantitatively in the setup described above, we apply IQPE to different initial product states.
In Fig.~\ref{fig:initstate}, we show a comparison between
(i) "Z" states that are initially polarized in the $Z$ basis, and
(ii) "ZX" states where initially half of the spins are polarized in the $Z$ and the other half in the $X$ basis (i.e., an
equal-weight superposition of $2^{L/2}$ product states in the $Z$ basis).
In Fig.~\ref{fig:initstate}, we observe that for the same $\Delta E/L$, the states obtained when
starting from "ZX" initial states have drastically larger entanglement entropy.

The reason for this can be understood in the picture of local conserved constants of
motion put forward for MBL states in Refs.~\onlinecite{Serbyn2013,Huse2013}: the eigenstates of the effective
Hamiltonian proposed there are simply product states in some fixed basis. Flipping a single spin in this basis will incur a large
energy penalty; however, by flipping many spins, one can obtain another product state which is exponentially
close in energy to the original state, but can be locally distinguished from the first state almost anywhere simply by measuring
in the preferred basis.
In the setup we consider, the local constants of motion are likely to be close to the physical $\sigma^z$ operators,
as the disorder is diagonal in this basis. The initial states polarized in the $Z$ basis thus differ from exact eigenstates
only by \emph{local} fluctuations, and thus have overlap with exact eigenstates that are far apart in energy.
In this case, IQPE is successful at isolating one or a few eigenstates with high accuracy.
The "ZX" states, on the other hand, can be thought of as superpositions of "Z" states that differ in $L/2$ spins,
and thus can have overlap with eigenstates that are nearby in energy. IQPE is thus less efficient at separating
these states, and for a given energy standard deviation the final state is a superposition of many nearby eigenstates.

\begin{figure}
  \includegraphics{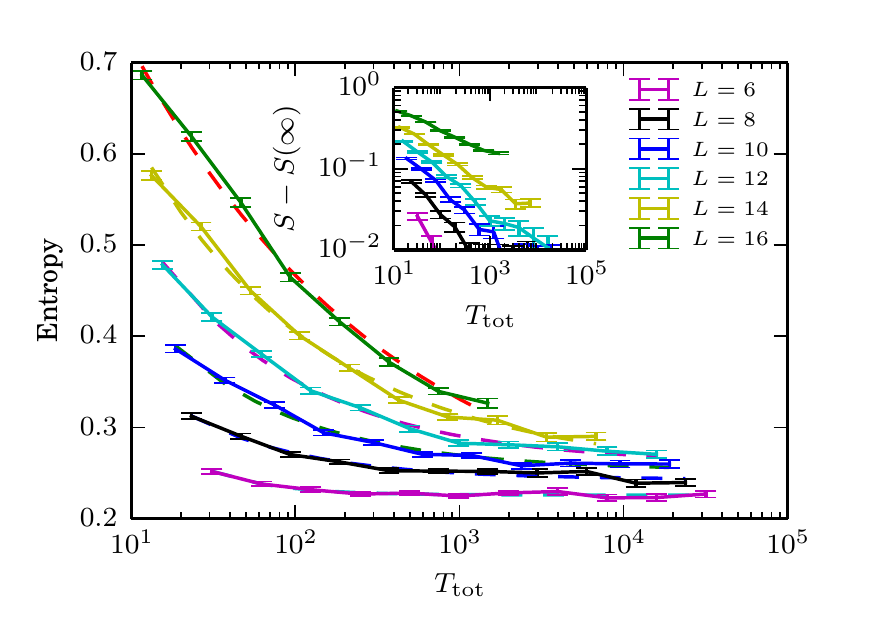}
  \caption{(Color online) Entanglement entropy at the center of the system. Dashed lines indicate a fit to $S = S_0 + a \Tt^{-b}$, where $a$ and $b$ are fit parameters. {\it Insert}: $S(\Tt) - S(\Tt \rightarrow \infty)$, where $S(\infty)$ has been obtained from the dashed fit in the main panel, on a double-logarithmic scale. \label{fig:S} }
\end{figure}

\begin{figure}
  \includegraphics{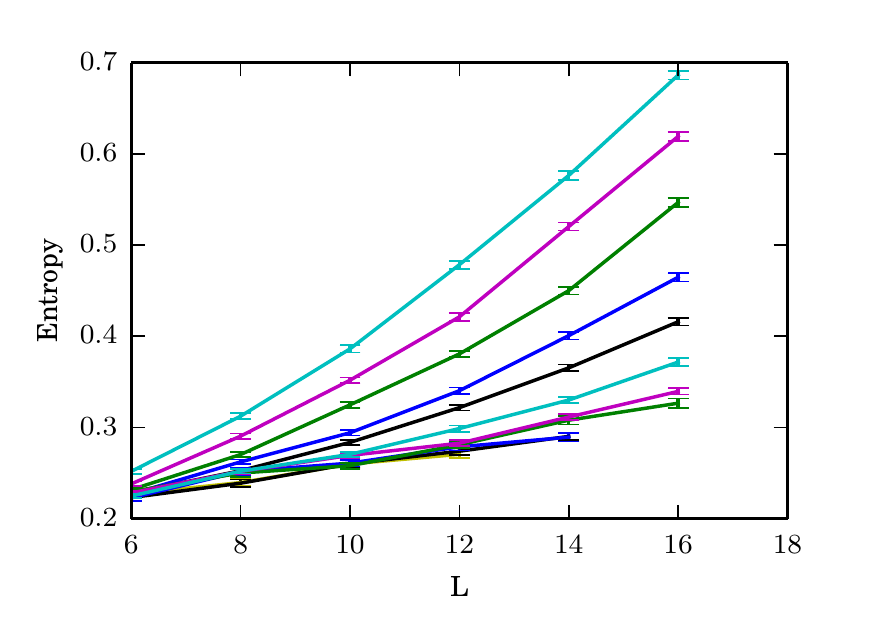}
  \caption{(Color online) The bipartite entanglement entropy for a center cut
as a function of system size $L$ in the strong disorder limit, as obtained by IQPE to accuracy (from top to bottom) $k = 10, 11, \ldots, 20$.
As the accuracy is increased, and the state begins to approach an energy eigenstate, the entanglement entropy decreases.
\label{fig:arealaw} }
\end{figure}

Concentrating on the "Z" initial states which, as argued above, give a better approximation to the exact
eigenstates for a fixed evolution time $\Tt$, we can ask how the entanglement entropy of
the final state depends on $\Tt$. Our results are shown in Fig.~\ref{fig:S}, where we show the entanglement
entropy for a cut in the middle of the system averaged over 10000 disorder realizations. We find good
agreement with a power-law fit, i.e. that the entanglement entropy approaches the exact value as $T^{-b}$
for some power $b$. In the inset of the figure, we show the same data after subtracting the constant term
on a double-logarithmic scale. For the small system sizes accessible to our simulations, the power $b$
appears to grow slowly as $L$ is increased. This dependence crucially affects how the time $\Tt$ necessary
to reach a given error in entropy $\varepsilon$ scales in the system size. Assuming that this growth is
sufficiently slow once an asymptotic regime is reached,
we can observe an area law by evolving for a time that is polynomial in the system size. To illustrate
this, we show in Fig.~\ref{fig:arealaw} the entropy vs system size $L$ for different choices of $\Tt$, and
observe that as $\Tt$ is increased an area law is observed for an increasing range of system sizes.

\section{Measurements} \label{sec:meas}

\subsection{General Remarks}

Suppose now that we have obtained an energy eigenstate $|{\psi_E}\rangle$
at some random energy $E$ by quantum phase estimation. What can we do with it to probe many-body localization?
As mentioned in the introduction, in contrast to classical simulations we cannot simply examine the wavefunction
to deduce its properties, and we are limited to performing unitary operations and projective measurements. 
Each such projective measurement yields at most $N$ binary numbers, where $N$ is the number of qubits
used for representing the physical system; the expectation value is then reconstructed by averaging over
many such measurements.

This gives rise to an additional complication when characterizing eigenstates obtained with the method described
above: Since this method does not allow us to target a specific eigenstate, we are unlikely to encounter the same eigenstate
more than once, and since we only measure $k < N$ bits of the energy we cannot uniquely identify the state by
its energy. We thus average \emph{simultaneously} over eigenstates in some energy window, where the width of the
energy window depends on $k$, and over measurement outcomes.
Notably, within the constraints of this setup, there is no known way to extract the entanglement entropy
of the resulting eigenstates.

Nevertheless, there are several powerful ways in which eigenstates can be characterized under these constraints.
In the following, we discuss the examples of transport in weakly perturbed eigenstates (either with a weak global
perturbation or a local perturbation), and how to adapt recent spin echo proposals~\cite{Serbyn14}.

If, on the other hand, we consider states obtained by performing global quenches on easily prepared initial states,
the preparation procedure becomes reproducible: we can prepare the same initial state over and over and apply
the time evolution for the same time interval $t$, and thereby prepare the same final state many times and perform
repeated measurements on this state.
Although the signatures of many-body localization are not as clear in this setup,
they are more easily obtained due to the possibility of repeated measurements on multiple copies of a state.

\subsection{Linear Response and Transport in Perturbed Energy Eigenstates}
\label{sec:transport}

It is instructive to briefly consider how an isolated system at fixed energy can be probed in an experiment.
One way is to couple the system to another, better-understood auxiliary system and see how it reacts.
In a sense, this is similar to coupling it to a heat bath, but in the limit in which the coupling is very small and can
be turned on locally, so that the auxiliary system can be used in a manner analogous to a thermometer.
In a many-body localized phase, we do not expect particles (or energy) to flow into the auxiliary system.

\begin{figure}
  \centering
  \includegraphics{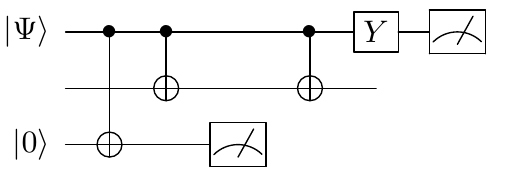}
  \caption{Quantum circuit to measure the current operator $i(c_i^\dagger c_j - c_j^\dagger c_i)$, as described in the main text.
  Here, the top two qubits are the qubits corresponding to the physical sites $i$ and $j$, and the bottom qubit is an ancilla qubit.
  \label{fig:cur} }
\end{figure}

A second possibility is to ``tilt'' the system. Having obtained an approximate eigenstate $|E\rangle$, we can evolve it for time $T$ with the Hamiltonian:
\begin{equation}
H = H_{\rm f} + \sum_j V j n_j
\end{equation}
where $H_{\rm f}$ is defined in Eq.~(\ref{eqn:H-fermions}).
This would correspond, in a cold atom experiment, to loading atoms into the trap with
fixed energy and then tilting the potential in the trap, as in Ref.~\onlinecite{Kondov13}.
For small $V$ and if the system is initially prepared close
to an eigenstate, this corresponds to a weak global perturbation.
We can then measure the current $i(c^\dagger_{j+1} c_j - \text{h.c.})$
at various locations within the system. Restricting to one dimension, where no Jordan-Wigner
strings need to be accounted for, this measurement is performed straightforwardly as shown in Fig.~\ref{fig:cur}.
Consider the following input state with an ancilla in state $\ket{0}$:
\begin{equation}
(a \ket{00} + b \ket{01} + c \ket{10} + d \ket{11})\ket{0}
\end{equation}
The first two qubits are the data qubits, which are the occupation numbers of sites $j$, $j+1$ and the third qubit
is the ancilla. The circuit in Fig.~\ref{fig:cur} implements the folllowing operations.
\begin{align}
\text{2 CNOTs:  } & (a \ket{00}+d \ket{11})\ket{0} + (b \ket{01} + c \ket{10})\ket{1} \\
\text{Measure 1:  } & b \ket{01} + c \ket{10} \\
\text{CNOT:  } & (b \ket{0} + c \ket{1} )\ket{1}
\end{align}
Here, in the step denoted as "Measure 1", we measure the ancilla qubit. If the measurement outcome is 0, the total measurement is 0 and the remaining steps need not be performed. If the outcome is $1$, then
CNOT is performed on the data qubits, followed a measurent of the first qubit in the $Y$ basis.

If the system is in a metallic state, then we expect the current to grow at short
times until it reaches a steady-state value, at a time on the order of the mean-free time $\tau \sim t/{W^2}$.
This value persists until the boundaries of the system reflect the current, at a time
$T\sim L^2/D \sim L^2 \frac{W^2}{t^3}$,
where $D$ is the diffusion constant, $D\sim {t^3}/{W^2}$. (We are working here in units in which the lattice spacing is $1$.)
Therefore, so long as $L\gg (t/W)^2$, there is a large interval of times over which the steady-state value can be observed.
In an MBL state, on the other hand, we expect the current to grow at short times, and then
rebound at a time on the order of $T\sim {\xi^2}/D$, where $\xi$ is the localization length. After that,
it will undergo damped oscillations, before reaching a vanishing steady-state value. Therefore, apart from a short time
transient, a current oscillating about and tending to zero will be observed.

Alternatively, we can study a "local quench": starting in an energy eigenstate, we can perturb the system locally,
e.g. by flipping the spins in
a small region $R$ (possibly even a single spin), and study the resulting time evolution.
Unlike a global quench, which is expected to show unbounded
growth of the entanglement, a local quench of an MBL state is expected to disturb the
system only in a localized region. This can be traced by following the spreading of correlations in the system
and observing, e.g., a zero-velocity Lieb-Robinson bound~\cite{Hamza2012}. Alternatively, 
we can measure the energy current at distant locations.
The energy current (as per Noether's theorem) on the link between $j$ and $j+1$ is: $i(c^\dagger_{j+2} c_j - \text{h.c.}) + i(c^\dagger_{j+1} c_{j-1} - \text{h.c.})$. This vanishes exponentially with distance at long times
in the MBL phase.

\subsection{Spin-Echo}
\label{sec:spin-echo}

One interesting variant on a local quench was suggested by Serbyn et al.~\cite{Serbyn14}.
This is most easily described in the context of a spin model (\ref{eqn:H-spins}).
In their proposal, one begins with a system in an initial product state in the $\sigma^z$ basis, except for
the $i^{\rm th}$ spin, which is in a ${\sigma^x_i} = +\frac{1}{2}$ eigenstate. The system is evolved for time $T$. The
$i^{\rm th}$ spin is then reversed by applying $\sigma^x$ and the system is evolved again for time $T$.
If the Hamiltonian were diagonal in the $\sigma^z$ basis, this would return the $i^{\rm th}$ spin
to its initial state. If spin-flip terms do not change the physics qualitatively, i.e. if the system is
adiabatically-connected to one in which the Hamiltonian is diagonal in the $\sigma^z$ basis, then
the $i^{\rm th}$ spin will not return precisely to its initial state, but to a state with
$\langle \sigma^x_i \rangle > 0$. Since this does not distinguish between many-body localization
and single-particle localization, Serbyn et al. \cite{Serbyn14} propose that some other spin or set of spins
in a region $R$ far from $i$ is manipulated (e.g. with a $\pi/2$ pulse). In a non-interacting localized state, this would have
no additional effect due the absence of coupling between spins in region $R$ and the $i^{\rm th}$ spin.
In a many-body localized state, however, there would be an intermediate range of $T$ values over which
$\langle \sigma^x_i \rangle$ would show power-law decay before saturating to a non-zero value at large $T$.
In a delocalized state, by contrast, $\langle \sigma^x_i \rangle$ decays rapidly to zero with $T$.
It is a straightforward matter to stop the time-evolution of a quantum state to reverse spin $i$ and to manipulate
spins in Region $R$ before continuing the time evolution.

\section{Errors}

Thus far, we have assumed that our quantum computer has infinite coherence time and that all operations can be performed flawlessly.
Obviously, this will not be the case, and errors must be taken into account in any appraisal of the prospects for applying a quantum computer.
One type of error that may be relatively benign is Trotterization errors. Such errors, which are systematic, may be re-interpreted as
a modification of the Hamiltonian. This effective Hamiltonian will have non-local terms induced by higher-order commutators of the original
Hamiltonian terms. However, such commutators are exponentially suppressed in their order, and thus the effective Hamiltonian has,
at worst, exponentially decaying terms. If we are interested in universal properties of MBL states,
such a modification of the Hamiltonian will be unimportant.

Much more serious errors are caused by the environment. However, of these, bit flip errors may be far more problematic than phase errors.
Suppose that the Hamiltonian for an MBL state can be written in terms of quasi-local conserved quantities $\tau^z_i$~\cite{Serbyn2013,Huse2013}:
\begin{equation}
H = \sum_i h_i \tau^z_i  + \sum_{i,j} J_{ij} \tau^z_i \tau^z_j + \ldots
\end{equation}
Then, the addition of a coupling to the environment that is diagonal in this basis (and, therefore, leads only to phase errors) takes the form:
\begin{equation}
H_{\rm sys-env} = \sum_i \tau^z_i B({X_a})
\end{equation}
Here, $B({X_a})$ is the effective field, which depends on the environment degrees of freedom, $X_a$, and,
therefore, entangles them with the quasi-local conserved quantities $\tau^z_i$.
However, such a coupling clearly has no effect on many-body localization.

Bit-flip errors, on the other hand, may have a rather drastic effect, in particular during the final stages of the quantum phase estimation: once the state is close to an eigenstate, a flip of a single spin will change the energy of the state by a large amount. Due to its recursive nature, this will lead to failure of the IQPE algorithm, and bit flip errors must thus be corrected at a lower level. This may be achieved by using topological qubits~\cite{Nayak08}, or by performing some error correction on the physical qubits. If we only take care to correct bit-flip errors, then we can essentially use a classical error correcting code. Consider, for illustrative purposes, a $[7,4]$ Hamming code. If we assume flawless error detection and recovery, then an initial error rate of $\varepsilon$ is lowered to an error rate of $21 \varepsilon^2$. In order to perform $10^6$ gates, we would, thus, need an gate fidelity of $99.99\%$,
at a cost of encoding $4$ sites in $7$ qubits, corresponding to a less than two-fold increase
in the required number of qubits.

\section{Conclusion}

Since classical computers are limited, for the foreseeable future, to the study of MBL systems
of approximately $20$ sites or less, a quantum computer need not be very large to accomodate
a significantly larger system. From the preceding considerations, it appears that a system of $50$ sites could
be simulated with fewer than 100 physical qubits, assuming realistic bit-flip error rates.
Moreover, as we have shown, a quantum computer can, in a
straightforward manner, manipulate such a system in ways that would be very time-consuming
with a classical computer. Two important features of many-body localization pave the way
to the practical application of a quantum computer:
(i) The Hamiltonian is local and can be written in terms of just 3 groups of non-commuting terms.
This greatly reduces the number of gates required for time evolution.
(ii) Since we are interested in dynamical properties and properties of highly excited states, possibly
close to a phase transition where the states acquire a volume law, classical computers are limited
to very small systems on the order of 20 sites.
Even deep in the localized phase, where states exhibit an area law and efficient algorithms exist to
find the ground state, there are no presently known algorithms that efficiently find highly excited states.
(iii) We are primarily interested in universal features and can, therefore, tolerate certain types of errors,
unlike in the case of Shor's algorithm or applications to quantum chemistry.
As a result a relatively small quantum computer can, in a reasonable time, evolve an initial state to longer times $t$
than would be possible with a classical simulation. In addition, a quantum computer can be used to
find an approximation to a typical energy eigenstate of a Hamiltonian.
Both applications of a quantum computer can give insights into many-body localization.

It is worth emphasizing that these results can be complementary to those obtained with a classical computer. We do not
have access to a classical representation of a quantum state prepared with a quantum computer. Moreover, we cannot prepare multiple
copies of the same approximate energy eigenstate. For these reasons, there is no obvious way to compute the entanglement entropy
of an approximate energy eigenstate. However, we can study features of MBL systems that would be very difficult with a classical computer, such
as transport, spin echo effects, and the long-time approach to a thermal or non-thermal state.

If one regards the quantum computation as a highly idealized model for an experiment on an isolated
quantum system, our results imply that the properties of eigenstates can be observed in the laboratory
with resources polynomial in the system size. This is a non-trivial insight, as exact eigenstates cannot
generally be prepared to very high accuracy unless exponentially large resources are used. This gives
additional justification for studying the properties of a many-body localized phase in its energy eigenstates,
but also has implications for experiments.

Indeed, some of the proposals we discuss bear great similarity to experimental approaches for example
in cold atoms systems.
For example, the transport scenario of measuring the response of an energy eigenstate to a weak tilt is relevant to experiments such
as the one reported in Ref.~\onlinecite{Kondov13}.
In this experiment, fermionic atoms are loaded into an optical lattice in a trap. A speckle pattern disorders the
potential in the trap. The atoms carry spin-$1/2$
and interact via an on-site Hubbard-like interaction. In addition, the atoms are not in their ground state, but have some energy
density that is fixed when they are loaded into the trap. Both of these conditions indicate that this system is in a regime
in which many-body localization could be observed. The trap is tilted and then the momentum distribution is measured.
When the disorder is weak, the momentum distribution is skewed by the tilt. When it is strong, the momentum distribution is
unaffected. This is broadly consistent with many-body localization, but it is difficult to to distinguish a transition from a
crossover and, therefore, difficult to determine whether the putative localized phase in the experiment is, in fact, a metallic phase
with small but non-zero conductivity.

\acknowledgements

We acknowledge useful discussions with Jon Yard and Rahul Nandkishore.
Simulations were performed using the ALPS libraries~\cite{Bauer2011-alps}.

\bibliography{qc4mbl}

\end{document}